\documentclass{aa}
\usepackage{graphicx}
\usepackage{txfonts}
\begin{document}
   \title{Jet emission in \object{NGC\,1052} at radio, optical, and X-ray frequencies}

   \author{M.\,Kadler
          \inst{1}
          \and
          J.\,Kerp\inst{2}
          \and
	  E.\,Ros\inst{1}
	  \and
          H.\,Falcke\inst{1}\thanks{\emph{Present Adress:} Radio Observatory, ASTRON, P.O. Box 2, 7990 AA Dwingeloo, The Netherlands}
          \and
          R.\,W.\,Pogge\inst{3}
          \and
	  J.\,A.\,Zensus\inst{1}
          }

   \offprints{M. Kadler, \email{mkadler@mpifr-bonn.mpg.de}}

   \institute{Max-Planck-Institut f\"ur Radioastronomie Bonn,
              Auf dem H\"ugel\,69, 53121 Bonn, Germany\\
         \and
              Radioastronomisches Institut, Universit\"at Bonn,
	      Auf dem H\"ugel\,71, 53121 Bonn, Germany\\
         \and
	      Department of Astronomy, Ohio State University, 
	      140 West 18th Avenue, Columbus, OH 43210-1173, U.S.A.}

   \date{Received 28 July 2003 / Accepted 19 February 2004}

   \abstract{{ We present a combined radio, optical, and X-ray study of the nearby LINER galaxy
NGC\,1052.
Data from a short 
(2.3\,ksec) {\it CHANDRA} observation of NGC\,1052 
reveal the presence of various jet-related X-ray emitting regions, 
a bright compact core and unresolved knots in the jet structure as well as
an extended emitting region inside the galaxy well aligned with
the radio synchrotron jet-emission.
The spectrum of the extended X-ray emission can best be fitted with a
thermal model with $kT = (0.4-0.5)$\,keV, while the compact core exhibits
a very flat spectrum, best approximated by an absorbed power-law with
$N_{\rm H} = (0.6-0.8) \times 10^{22}\,{\rm cm^{-2}}$. 
We compare the radio structure to an optical ``structure map'' from a
{\it Hubble Space Telescope} ({\it HST}) observation and find a good
positional correlation between the radio jet and the optical emission cone.
Bright, compact knots in the jet structure are visible in all three
frequency bands whose spectrum is inconsistent with synchrotron emission.}
  } 

   \maketitle
%

\section{Introduction}
NGC\,1052 is a nearby\footnote{$D = $ 22.6\,Mpc (assuming $z =$ 0.0049
(Knapp et al. \cite{Kna78}) and $H_0 = 65$\,km\,s$^{-1}$\,Mpc$^{-1}$). At this distance
1\,arcsec corresponds to $\sim$110\,pc.} elliptical galaxy 
which harbors a low-luminosity
active galactic nucleus (LLAGN) in its very center (L$_{1-100{\rm GHz}}=4.4 
\times 10^{40}$\,erg\,s$^{-1}$; Wrobel \cite{Wro84}).
{ It hosts} a two-sided radio jet 
emanating from the nucleus and reaching out to { kiloparsec-scales} which
is, however, still fully enclosed within the stellar body of the optical
galaxy.
In the optical, the spectrum of NGC\,1052 is characterized by strong forbidden
lines from low-ionization states which has made NGC\,1052 the prototypical
LINER (low-ionization nuclear emission line region; Heckman \cite{Hec80}) galaxy.
As for LINERs in general, it has long been argued whether these low-ionization
lines in NGC\,1052 are excited by a central photo-ionizing source (e.g., Gabel et al. 
\cite{Gab00}) or if shock heating is the dominant mechanism (e.g., Sugai \& Malkan
\cite{Sug00}). While there is overwhelming evidence for the 
presence of an active galactic nucleus (AGN) in NGC\,1052,
the role of shocks in this galaxy is still unclear. 
The improved angular resolution in X-rays offered by {\it CHANDRA} makes it possible 
for the first time to image the distribution of X-ray emission on
the same scales as accomplished by connected radio 
interferometers, e.g., MERLIN (Multi-Element Radio-Linked Interferometer
Network).
Disentangling the contributions of compact 
(i.e., $<$1\,arcsec) nuclear and extended (i.e., $>$1\,arcsec) X-ray
emitting regions to the total amount of X-ray emission of NGC\,1052
can serve as an important tool to study the interaction between the
radio jet plasma and the ambient interstellar medium. 

NGC\,1052 has been observed by all major X-ray missions of the pre--{\it CHANDRA} era, like
{\it Einstein} (Mc Dowell \cite{McD94}), {\it ASCA} and {\it ROSAT} (Weaver et al. \cite{Wea99}), and 
{\it Beppo Sax} (Guainazzi \& Antonelli \cite{Gua99}).
For these X-ray missions, NGC\,1052 appeared as a point-like
X-ray source. { The X-ray spectrum of NGC\,1052 
is extremely flat, a finding that led to the proposal
of an advection dominated accretion flow (ADAF) as the origin of the 
observed X-ray
emission (Guainazzi et al. \cite{Gua00}).} 
To model the AGN X-ray spectrum above $E \simeq $ 2\,keV absorbing column
densities in excess of $10^{23}$\,cm$^{-2}$ have been discussed, supporting 
the idea of a high density obscuring torus, comparable to column 
densities found in other AGN (e.g., Malizia et al. \cite{mali97}, Risaliti et al. \cite{risa02}).
Independent evidence for the existence of an obscuring torus at the center of
NGC\,1052 is obtained from Very Long Baseline Interferometry (VLBI) observations
in the radio regime: 
on { parsec-scales} NGC\,1052 exhibits a twin jet structure with a prominent
emission gap between both jets (see e.g., Kadler et al. \cite{Kad03b}).
The inner part of the western jet shows 
a strongly inverted radio spectrum, which was first discovered
by Kellermann et al. (\cite{Kel99}) (see also Kameno et al. \cite{Kam01}). The cm-wavelength spectral 
index in this central region is larger than 2.5, exceeding the
theoretical limit for synchrotron self-absorption. 

Combined studies of the core region of NGC\,1052 in the radio and X-ray regime
are of essential importance { for 
constraining the physical properties of the { parsec-scale} radio jet and the
obscuring torus
as well as to determine the nature of the nuclear X-ray emission.}
In this paper we present a combined radio, optical, and X-ray study
of the jet-related emission in NGC\,1052 on arcsecond scales.
In particular, we focus on the soft X-ray excess in the source-spectrum 
below $E = $ 2\,keV.
This soft component was identified first by Weaver et al. (\cite{Wea99}) based on {\it ROSAT}
PSPC data. {\it CHANDRAs} superior angular resolution makes it possible 
to present evidence 
that this soft excess emission is 
associated with the well known radio jet.

In Sect.~\ref{obs} we present the {\it CHANDRA}, MERLIN, and 
{\it HST} data as well as their reduction.
In Sect.~\ref{multi-structure} we discuss the arcsecond-scale
morphology of NGC\,1052
in the radio, optical, and X-ray regime and the correlations between the
different wave bands.
In Sect.~\ref{spectroscopy} we derive models for the nuclear and
extended X-ray emission and
Sect.~\ref{summ} summarizes our conclusions.

\section{Observations and data reduction}
\label{obs}
\subsection{{\it CHANDRA} data}
\begin{figure}[t!]
   \centering
\includegraphics[clip, width=\linewidth]{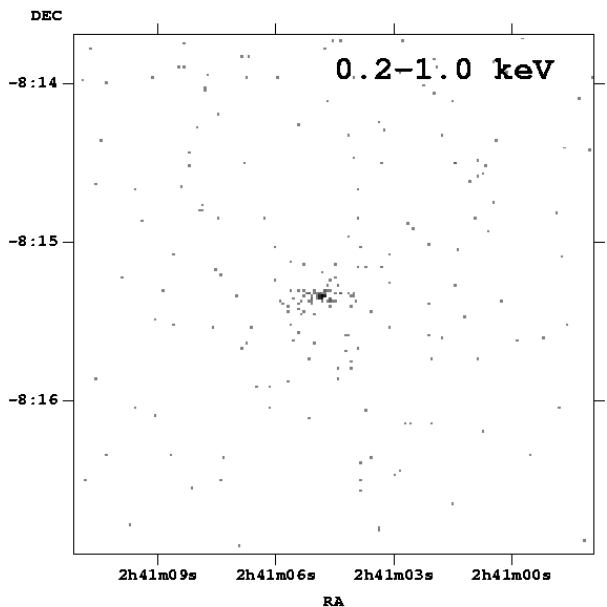}
{\caption[]{\small Raw {\it CHANDRA} image of NGC\,1052 showing the spatial distribution of
photon events in the energy range ($0.2-1.0$)\,keV.
\label{fig:events}}}
\end{figure}
{\it CHANDRA} observed NGC\,1052 on August 29/30, 2000\footnote{The {\it CHANDRA} data were
taken from the public archive ({\tt http://cxc.harvard.edu/cda/chaser.html})
and analyzed using standard methods { within the software package {\sc ciao 3.0}
using the calibration data base {\sc caldb 2.2.5} version.}}.
During the 2342\,s observation, the Advanced CCD Imaging Spectrometer
(ACIS) Chip S3 was in the focus of the High Resolution Mirror Assembly (HRMA).
The ACIS-S3 detector offers high angular resolution as well as information on
the X-ray source spectrum because of its intrinsic energy resolution
and has a higher sensitivity in the soft X-ray energy regime below
$E\,<\,1$\,keV then the front-side illuminated detectors of 
{\it CHANDRA}.

The nucleus of NGC\,1052 is the brightest X-ray source within the field of
interest.
The count rate of 0.12\,${\rm cts\,s^{-1}}$ is sufficiently high to affect 
the measured 
AGN X-ray spectrum by the pile-up effect.
The pile-up effect changes the shape of the measured X-ray spectrum as well as
the measured count rate, because during a single read-out
period of the chip multiple X-ray photons may be detected within a single
pixel. Due to insufficient time resolution their combined signal is
registered as a single photon event.
The presented {\it CHANDRA} observation was performed using the standard
timed exposure mode. 
Depending on the detailed shape of the AGN X-ray spectrum up to
27\% of the available data can be affected by the pile-up effect.
In principle, this introduces a bias which results in spectral
hardening and mimics lower count rates.

{ The spatial distribution of photon events on the ACIS-S3 chip during
the observation of NGC\,1052 is shown in Fig.~\ref{fig:events} for the
energy range ($0.2-1$)\,keV. Even from
the small number of counts during this snapshot observation the distribution
of soft X-rays clearly differs from a point spread function.}
Using the {\sc csmooth} program which is part of the {\sc ciao} software
we also produced
an adaptively smoothed map of the field of interest { (see Fig.~\ref{fig:truecolor}) 
for which we reduced} the angular
resolution of the data from 0.5\,arcsec to 4\,arcsec and set a minimum
significance threshold to 3$\sigma$.

\subsection{MERLIN data}
A MERLIN observation of 
NGC\,1052 at 1.4\,GHz was { performed} on November 22, 1995. The data from this
experiment have been obtained from the public archive\footnote{
{\tt http://www.merlin.ac.uk}} and analyzed applying standard
methods using the program {\sc difmap}. Two different maps of the
brightness distribution of NGC\,1052 at 1.4\,GHz were produced. First,
a strong $(u,v)$-taper was used to map the extended
emission resulting in a restoring beam of ($1.5 \times 1.1$)\,arcsec
at a position angle (P.A.) of $-31^\circ$. Second, a pure naturally weighted
image was produced yielding a restoring beam of 
($0.5 \times 0.3$)\,arcsec at a P.A. of $27^\circ$.

\subsection{{\it HST} data}
For this paper, we re-examined the archival
{\it HST} 
data of NGC\,1052 previously published by
Pogge et al. (2000).  The archival data are three narrow-band F658N
filter images acquired with the WFPC2 in the PC1 detector (pixels
scale $\sim0.046$\,arcsec) with integration times of 200, 800, and
900\,s, respectively.  Cosmic rays were cleaned from the individual
images using a version of the L.A.Cosmic Laplacian edge-detection
algorithm described by van Dokkum (\cite{VanD01}).  { The
cleaned} images were added together after
registration using a field star in the image to remove the
inter-image offsets.  To enhance the faint emission features noted by
Pogge et al. (\cite{Pog00}), we created a ``structure map'' of the
combined, cleaned images following the technique described by Pogge \&
Martini (\cite{Pog02}). We used a model { point spread function}
for the F658N filter and PC1
camera
generated using the {\sc TinyTim} package (Krist \& Hook \cite{Kri99}).  Structure
mapping is an improved method of image contrast enhancement
(compared to traditional unsharp masking).
In brief, the technique suppresses
the large-scale starlight distribution, enhancing faint emission and
absorption (e.g., dust extinction) features in the image.
   \begin{figure*}[tbh]
\vbox{\includegraphics[clip, width=12cm]{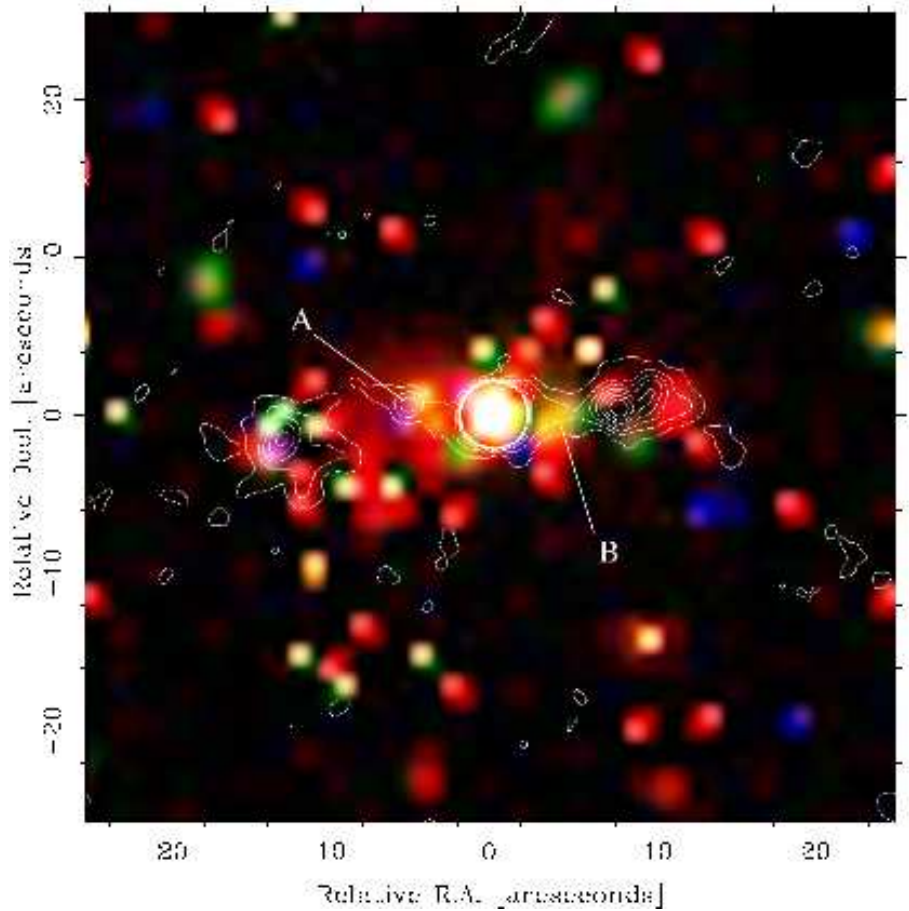}}\vspace{-6cm}
\hfill \parbox[b]{5.5cm}
{\caption{\small Three color {\it CHANDRA} image of the { jet-associated X-ray emission} in NGC\,1052
with the 1.4\,GHz MERLIN radio image overlaid in contours. The {\it CHANDRA}
image has been
smoothed to a resolution of 4\,$^{\prime\prime}$ to increase sensitivity.
Red represents photons between 0.1\,keV and 1\,keV, green between 1\,keV and 2\,keV
and blue ($2-3$)\/keV. The clipping level is 3$\sigma$.
The MERLIN image was
restored with a circular beam of 1.7\,arcsec FWHM.
Contours are shown for
{ (0.5, 1.5, 2.5, 3.5, 4.5, 5.5, 6.5) $\times$ 1\,mJy/beam.}
The { jet-associated} emission is dominated by soft X-rays below 1\,keV,
whereas the AGN appears white, representing a rather flat X-ray spectrum.
              \label{fig:truecolor}}}
    \end{figure*}

\section{The multi-waveband jet structure}
\label{multi-structure}
\subsection{ Jet-associated X-ray emission in NGC\,1052}
The {\it CHANDRA} image of NGC\,1052 
{ in the energy range ($0.3-3.0$)\,keV}
is shown in Fig.~\ref{fig:truecolor}.
Strong X-ray emission from the AGN is seen in this image, as well as diffuse
extended emission well aligned with the radio jet, whose MERLIN
image is superimposed in
contours.
In Sect.~\ref{jetspec} and Sect.~\ref{agnspec} we discuss the X-ray spectra
in detail and present evidence that the soft excess component found previously
in the {\it ROSAT} spectra of the nucleus (Weaver et al. \cite{Wea99}) is associated with
this soft, extended { jet-associated} emission.

To the first order,
the radio and the X-ray jet are aligned on arcsecond scales 
and their extent appears to be the same.
However, the intensity of synchrotron emission in the radio lobes 
appears to be anti-correlated
with the soft X-ray intensity distribution in so far as the radio
hotspots on both sides correspond to regions on the ACIS-S3 chip
which did not detect any photon events. 
East and west of the nucleus two relatively bright X-ray emitting regions
coincide with emission knots in the radio regime.
Bright optical emission knots are also present in these areas 
(Pogge et al. \cite{Pog00}).

\subsection{Radio emission on arcsecond scales}
\begin{figure}[htb]
\centering
\includegraphics[clip, width=8.5cm]{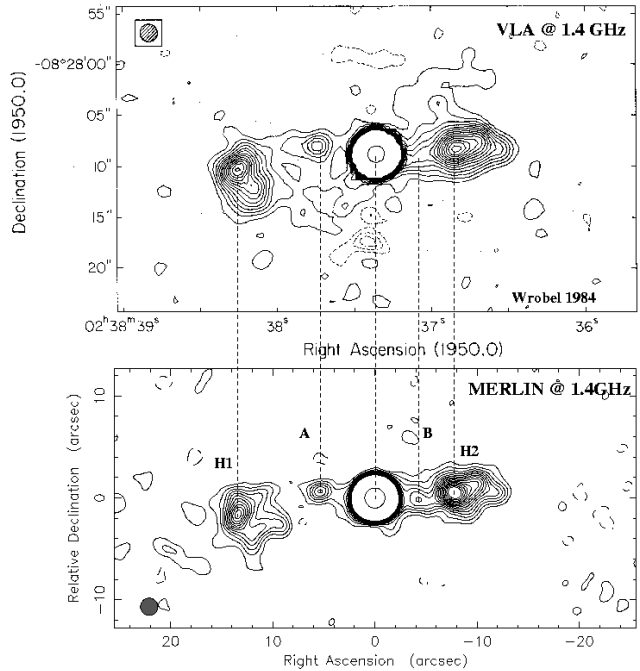}
\caption{\small Comparison
between the VLA image of NGC\,1052 observed in 1980 (Wrobel \cite{Wro84}) and the MERLIN image from
1995, both at 1.4\,GHz. Both maps have been convolved with the same (circular) restoring beam
of 1.7\,arcsec FWHM. Contours at ($-$1, 1, 2, 3, 4, 5, 6, 7, 8, 9, 10, 11, 12, 13, 14, 15,
800) $\times$ 0.5\,mJy are given in both maps (the VLA map also shows negative
contours at ($-4$, $-3$, $-2$) $\times$ 0.5\,mJy). The hot spots (labeled as H1 and H2)
do not show any { positional changes.}
The emission peak of the eastern knot (A)
is about $0.5$\,arcsec closer to the core in 1995.
              \label{fig:compare}}
    \end{figure}

The large-scale radio structure of NGC\,1052 visible in the tapered MERLIN image 
reveals some differences to the 1980 VLA image of 
Wrobel (\cite{Wro84}) (see Fig.~\ref{fig:compare}).
The VLA, which was operated in its A configuration during the 1980 observations, 
was more sensitive to extended structures, which partially have been
resolved out by MERLIN.
The core has varied, with a flux density of $\sim$0.74\,Jy in December 1980 
and $\sim$1.01\,Jy in November 1995. The western hot spot has increased in
flux density. A knot in the western jet (labeled as B { in the MERLIN image}) 
was not visible as a local maximum in the 1980 VLA image.
By aligning the core positions at both epochs,
we found that the eastern knot (labeled as A),
has moved about $0.5$\,arcsec inwards. However, an  examination of the
pure naturally weighted MERLIN image (see Fig.~\ref{fig:structure_map}) reveals that
knot A is composed of two isolated sub-components (A\,1 and A\,2)
separated by the same distance of $\sim$0.5\,arcsec.
This suggests that the inner sub-component might have increased
in flux density between 1980 and 1995, causing an apparent shift of the 
blended knot structure
in Fig.~\ref{fig:truecolor}. This offers a more plausible explanation than an inward motion 
with a velocity of $\sim$10\,$c$.

\subsection{Optical structure}
\begin{figure*}[htb]
\vbox{\includegraphics[clip, width=12cm]{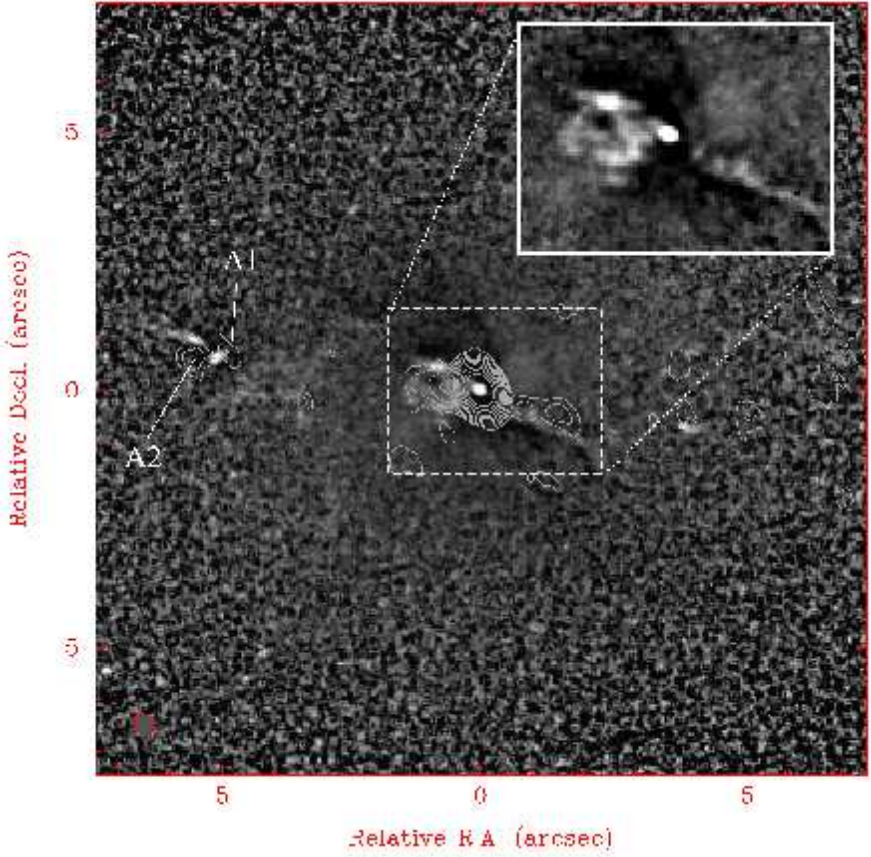}}\vspace{-4.5cm}
\hfill \parbox[b]{5.5cm}
{\caption[]{\small {\it HST} structure map of NGC\,1052. Dark regions represent dust obscuration while
bright regions are locations of enhanced emission. Overlaid is the pure naturally
weighted MERLIN map showing the core of NGC\,1052 with sub-arcsecond resolution. Contours
at ($-$1, 1, 2, 4, 8, 16, 32, 64, 128, 256, 512, 1024) $\times $ 0.5\,mJy/beam
are shown. The inlaid panel shows an enlarged view of the marked box around the
area of the optical emission cone.
\label{fig:structure_map}}}
\end{figure*}
The structure map of NGC\,1052 is shown in Fig.~\ref{fig:structure_map} with the radio contours
of the untapered, full-resolution MERLIN image superimposed.
The jet--counterjet structure is aligned with the optical emission
cone. The dark band perpendicular to the radio jet might be an artifact of the
image processing. Two optical emission knots are located at the
edges of the two radio sub-components of knot A. The optical emission knot in the
west coincides roughly with a weak ($\sim 1 \sigma$) radio feature while the
stronger radio knot 1.5\,arcsec further out has no corresponding bright
optical counterpart.
The origin of the optical emission remains unclear since there is no
continuum image to subtract from the H$\alpha$
filter image of NGC\,1052. 
The conical morphology in this LINER 1.9 galaxy is remarkably
similar
to the structure of the narrow-emission-line region typically observed 
in Seyfert\,2 galaxies (e.g., Falcke et al. \cite{Fal98}).
This suggests that the optical emission cone 
is due to line emission rather than continuum emission. 
The optical flux density of the two eastern knots of ($68\pm1$)\,$\mu$Jy
exceeds the power-law extrapolation from the radio to the X-ray regime by
almost three orders of magnitude. This 
is a strong argument  against the synchrotron emission process, 
but compatible with relatively strong line emission exceeding the
contimuum emission of the knots in this narrow band.

\section{X-ray imaging spectroscopy}
\label{spectroscopy}
Using {\it CHANDRA's} high angular resolution, it is possible to obtain
separate spectra of
the nucleus and the { diffuse jet-associated X-ray emission.}
We selected an annulus centered at the nucleus position which excludes the
nucleus itself but includes the whole area of the diffuse emission. 
To avoid any
contamination by the nuclear X-ray emission, a diameter of 
3\,arcsec for the inner annulus
ring was used.
The nuclear X-ray spectrum was extracted up to a diameter of 2\,arcsec.
The extracted X-ray data of both regions were corrected for the unrelated X-ray
background emission, using a region located close by
without significant point sources. We also applied a correction for the quantum
efficience degeneracy of the ACIS detector.

\begin{figure}[htb]
   \centering
   \includegraphics[clip,width=\linewidth]{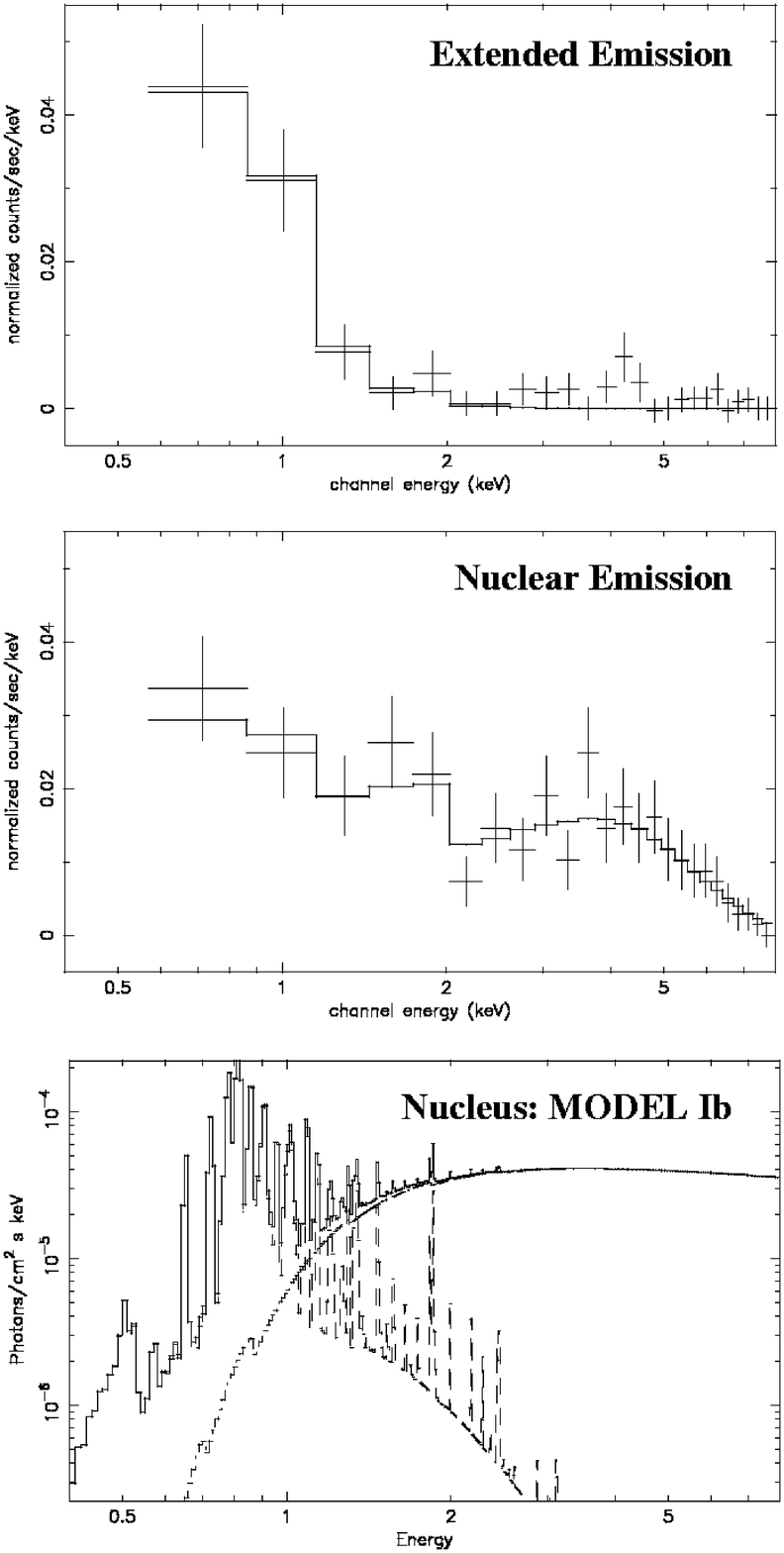}
      \caption{\small Top panel: X-ray spectrum of the extended X-ray jet emission and the Raymond/Smith plasma model
folded with the detector response matrix (solid line). Middle panel: X-ray spectrum of the nuclear 
emission and the best fit of model Ib (solid line, compare Table~\ref{agnspec}). 
Bottom panel: The unfolded model Ib spectrum representing an absorbed Raymond/Smith plasma
whose model parameters have been held fixed at the best fitting values to the extended X-ray 
emission spectrum (see Table \ref{tab:extended}) plus a power-law with an additional absorber.
         \label{fig:specs}}
   \end{figure}

Fig.~\ref{fig:specs} reveals the much softer  
spectrum of the diffuse X-ray emission compared to the nuclear spectrum.
The bulk emission originates below $E\,<\,2$\,keV,
while the X-ray spectrum of the nucleus has an additional hard X-ray component.
Remarkable is the soft X-ray emission below $E\,<\,2$\,keV towards the nucleus.
{\it ASCA}, as well as the {\it ROSAT} PSPC, could not separate the nucleus and the
jet spatially, but a soft excess X-ray emission had already been
detected in the X-ray spectrum from the PSPC data.

\subsection{The X-ray spectrum of the extended emission}
\label{jetspec}
\begin{table*}[htb]
      \caption[]{\small Best-fitting parameters for the spectral fits to the extended { jet-associated} emission
         \label{tab:extended}}
\centering
            \begin{tabular*}{13cm}{ccccccc}
            \hline
            \noalign{\smallskip}
            Model$^{\mathrm{(a)}}$      &  $N^{\mathrm{(b)}}_{\rm H}$  & $kT$   &$\Gamma^{\mathrm{(c)}}$ & $F_{\rm abs}^{\mathrm{(d)}}$ & $F_{\rm unabs}^{\mathrm{(d)}}$ & $\chi^2_{\rm red}$ ($\chi^2$/d.o.f.)\\
                                      &  [10$^{22}$\,cm$^{-2}$]               & [keV] & & [erg\,s$^{-1}$\,cm$^{-2}$] & [erg\,s$^{-1}$\,cm$^{-2}$] &\\
            \noalign{\smallskip}
            \hline
            \noalign{\smallskip}
            (R\&S)$^{\rm abs}$  & 0.49$_{-0.06}^{+0.08}$ & $0.48_{-0.08}^{+0.08}$&  - & 0.8$\times 10^{-13}$ & 3.9$\times 10^{-13}$ & 0.49 (10.8/22)   \\
            (M\&K)$^{\rm abs}$  & 0.41$_{-0.06}^{+0.08}$ & $0.41_{-0.07}^{+0.09}$&  - & 0.7$\times 10^{-13}$ & 3.4$\times 10^{-13}$ & 0.53 (11.6/22)   \\
            (PL)$^{\rm abs}$    & 0.06$_{-0.02}^{+0.07}$  & - &  $3.8_{-0.7}^{+0.6}$  & 1.5$\times 10^{-13}$ & 2.0$\times 10^{-13}$ & 0.52 (11.5/22)   \\
            \noalign{\smallskip}
            \hline
\noalign{\smallskip}
\multicolumn{7}{p{12.5cm}}{\scriptsize
$^{\mathrm{a}}$ R\&S: Raymond/Smith plasma, M\&K: Mewe/Kaastra plasma, PL: Power-law;
$^{\mathrm{b}}$ Absorbing column density; $2\times 10^{20}$\,cm$^{-2}$ is set as a lower limit;
$^{\mathrm{c}}$ Photon index;
$^{\mathrm{d}}$ (Un)Absorbed X-ray flux between 0.2\,keV and 8.0\,keV
}
\end{tabular*}
\end{table*}

To constrain the emission process of the diffuse extended emission, we fitted 
different models using {\sc xspec} (version 11.2):
the {\sc raymond} model corresponding to
a Raymond/Smith plasma (Raymond \& Smith \cite{Ray77}), the {\sc mekal} model describing 
a Mewe/Kaastra plasma (e.g., Kaastra \cite{Kas92}) and a simple power-law.
The results of the spectral fits are presented in Table \ref{tab:extended}.

Because the jet is located deep
inside the stellar body of the galaxy NGC\,1052, we have to determine 
the amount of photoelectric absorption distributed along the line of sight.
Using the thermal source models, we find an attenuating column
density value 
about one order of magnitude higher than the galactic
foreground column density belonging to the Milky Way 
(Hartmann \& Burton \cite{Har97}).
We attribute this additional X-ray attenuation to weakly ionized gas located
inside the galaxy itself.
The unabsorbed flux of the Raymond/Smith (Mewe/Kaastra) plasma in the range between
0.3 and 8\,keV  
corresponds to an intrinsic luminosity of 
{ $L=2.4(2.1) \times 10^{40}$\,erg\,s$^{-1}$} at the distance of the source
of 22.6\,Mpc.

Both plasma models fall below the measured count rate above 2.5\,keV (see Fig.~\ref{fig:specs})
suggesting some  
contribution of an additional hard X-ray component to the
spectrum of the extended emission.
Due to the low photon statistics and the large { uncertainties} 
we have not tried to account for this additional
spectral component by considering more complicated models.

Because of the low number of counts from the jet, we cannot 
unequivocally dismiss the power-law model, which gave an 
acceptable fit to the X-ray data.
However, we consider the power-law fit result of the X-ray jet { as unreliable
because of the unusually} steep photon 
index\footnote{The photon index $\Gamma$ is defined as the power-law
index of the spectrum given in units of [photons\,s$^{-1}$\,keV$^{-1}$].
It is related to the energy index $\alpha$, used in radio astronomy,
via: $\Gamma = 1 - \alpha$} (see
Table \ref{tab:extended}).


\subsection{The X-ray spectrum of the nucleus}
\label{agnspec}
Below { $E=2$\,keV} a soft excess is detected in the nuclear spectrum.
Assuming that a fraction of the diffuse { jet-associated} emission
originates within the immediate neighborhood of the nucleus,
we used a hybrid model for the X-ray spectral approximation of the 
observed intensity distribution.
\begin{equation}
I_{\rm obs} = (I_{\rm diffuse} + I_{\rm nucleus}\times e^{-\sigma\times N_{\rm H(torus)}})\times e^{-\sigma\times N_{\rm H(gal.)}}
\label{eq:i_obs}
\end{equation}
where $I_{\rm diffuse}$ is
the X-ray spectrum of the { diffuse jet-associated emission} 
and $I_{\rm nucleus}$ is the spectrum of the
nucleus.
Here, the photoelectric absorption produced by the torus is represented 
by $e^{-\sigma\times N_{\rm H(torus)}}$, while the X-ray absorption produced by the
interstellar medium of the elliptical galaxy is $e^{-\sigma\times N_{\rm H(gal.)}}$.

We assumed a power-law type X-ray spectrum of the central X-ray source
and modeled the soft component as a thermal plasma spectrum of the 
Raymond/Smith and Mewe/Kaastra type, (model I and II) respectively.
We did not try to fit for the metallicities and fixed these during the spectral
fitting procedure { given the poor statistics from the short integration time.} 
Model Ia(IIa) assumes solar abundances of the Raymond/Smith (Mewe/Kaastra)
plasma and allows the other parameters to be varied. In model Ib(IIb) the temperatures of the
plasma and the absorbing column density of the interstellar medium of the galaxy were fixed
to the best fitting values to the extended X-ray emission (see Table \ref{tab:extended}). 
In model Ic(IIc) the metal abundances were additionally changed to 25\% of the solar composition. 
Such low metal abundances typically occur only in dwarf galaxies but are 
found in NGC\,1052 from optical
line measurements of Sil$^\prime$chenko (\cite{Sil95}) who
reports values [Fe/H]$\sim -0.6$ for the nucleus and [Fe/H]$\sim -1$ for
the bulge of NGC\,1052, corresponding to metallicities of about 25\% and
10\%, respectively.
One way to accumulate a substantial amount of metal-poor
material is a merger or close encounter with an extremely
metal-poor dwarf galaxy. Evidence for such an event indeed exists from
multiple observational approaches (see Forbes et al. \cite{For01} and
references therein) in the case of NGC\,1052.

The statistically best fitting values are given in Table \ref{tab:coreparameter}. 
{ For all six models we derive a luminosity of $L=1.4 \times 10^{41}$\,erg\,s$^{-1}$.
Correcting for the effect of the two absorber model components increases the intrinsic
(unabsorbed) value to $L=(1.7-2.0) \times 10^{41}$\,erg\,s$^{-1}$.} 
The values of $N_{\rm H(gal.)}$ and $kT$ for the plasma component in the models Ia 
and IIa 
are very similar to the values determined for the extended plasma emission. 
For these models we derive { best fitting values for the} absorbing column density of 
{ $N_{\rm H(torus)} = (0.6-0.7) \times 10^{22}$\,cm$^{-2}$
as well as $\Gamma = 0.2-0.3$ for the  photon index. Similar} 
values for both parameters result from fixing $N_{\rm H(gal.)}$ and $kT$ as described above in
the models Ib and IIb. Reducing the metal abundances of the plasma component in model Ic(IIc)
forces the power-law to contribute more strongly to the soft part of the spectrum resulting in 
higher values of { $N_{\rm H(torus)} \sim 0.8 \times 10^{22}$\,cm$^{-2}$ and 
$\Gamma \sim 0.3$.} 
These values are still rather low. Data from other X-ray
observatories imply much higher values for  $N_{\rm H(torus)}$ as
well as for $\Gamma$ (Weaver et al. \cite{Wea99}; Guainazzi \& Antonelli 
\cite{Gua99}; Guainazzi et al.  \cite{Gua00}).
We note that the apparent discrepancy between 
previous X-ray observations and
the {\it CHANDRA} data
concerning the photon index might be due to the pile-up
degradation ($\sim 30$\,\%) of the latter.
The nuclear X-ray spectrum thus might appear
artificially flattened. 
The determined absorbing column density, however, is not so sensitive to the pile-up
effect, which mainly affects the
hard part of the spectrum while the amount of absorption is determined in
the soft X-ray regime.
The statistical quality of all six fits is very similar, with values of $\chi_{\rm red}^{2}$ ranging from 0.76 to 0.84,
and the low photon statistics (reflected in the values of $\chi_{\rm red}^{2} < 1$) do not allow us 
to dismiss any of these models. 

{ We note that a ``patchy'' absorber model in which a fraction of the central source is seen directly
and only a part of the source is covered by the absorber might be a more realistic model for
the nuclear X-ray emission of NGC\,1052, given the results of
Vermeulen et al. (\cite{Ver03}), who found that moving VLBI components in both jets show complex
light curves probably caused by substantial patchiness of the absorbing screen.
Additionally, they
find a complex H\,{\sc i} absorption line
spectrum with compact clouds of absorbing material at different relative velocities and different
locations along the parsec-scale jet structure.
The absorbing column density derived from ``patchy'' absorber models is 
expected to be considerably higher ($\sim 2 \times 10^{22}$\,cm$^{-2}$) than
the values derived from uniform absorber models.
However, given the low photon statistics, we omit the detailed discussion of such models with a larger 
number of free parameters.}

   \begin{table*}[htb]
      \caption[]{\small Best-fitting parameters for the spectral fits of the core emission
         \label{tab:coreparameter}}
\centering
\small{
\begin{tabular*}{16cm}{ccccccccc}
            \hline
            \noalign{\smallskip}

            Model$^{\mathrm{(a)}}$      &  $N^{\mathrm{(b)}}_{\rm H(gal.)}$  & $kT$ & $Z^{\mathrm{(c)}}$ & $N^{\mathrm{(d)}}_{\rm H(torus)}$  & $\Gamma^{\mathrm{(e)}}$  & $F_{\rm abs}^{\mathrm{(f)}}$ & $F_{\rm unabs}^{\mathrm{(g)}}$ & $\chi^2_{\rm red}$ ($\chi^2$/d.o.f.) \\
                                      & {[10$^{22}$\,cm$^{-2}$]}   &{[keV]}&   &{[10$^{22}$\,cm$^{-2}$]}     &   &  [erg\,s$^{-1}$\,cm$^{-2}$] & [erg\,s$^{-1}$\,cm$^{-2}$] &     \\
            \noalign{\smallskip}
            \hline
            \noalign{\smallskip}

Ia & $0.66^{+0.07}_{-0.07}$& $0.44^{+0.07}_{-0.06}$& 1$^{*}$& $0.74^{+0.56}_{-0.39}$ & $0.30^{+0.10}_{-0.08}$ & 2.3$\times 10^{-12}$ & 3.3$\times 10^{-12}$ & 0.84 (15.9/19)\\
Ib  & $0.49^{*}$& $0.48^{*}$& 1$^{*}$& $0.54^{+0.46}_{-0.34}$ & $0.20^{+0.10}_{-0.08}$ & 2.3$\times 10^{-12}$ & 2.9$\times 10^{-12}$ & 0.76 (16.0/21) \\
Ic  & $0.49^{*}$& $0.48^{*}$& 0.25$^{*}$& $0.84^{+0.55}_{-0.39}$ & $0.28^{+0.09}_{-0.09}$ & 2.3$\times 10^{-12}$ & 3.0$\times 10^{-12}$ & 0.74 (15.8/21) \\
\noalign{\smallskip}
IIa & $0.51^{+0.09}_{-0.04}$& $0.28^{+0.03}_{-0.04}$& 1$^{*}$& $0.45^{+0.59}_{-0.21}$ & $0.22^{+0.13}_{-0.05}$ & 2.5$\times 10^{-12}$ & 3.4$\times 10^{-12}$ & 0.89 (16.8/19) \\
IIb & $0.41^{*}$& $0.41^{*}$& 1$^{*}$& $0.62^{+0.46}_{-0.37}$ & $0.23^{+0.06}_{-0.09}$ & 2.3$\times 10^{-12}$ & 2.8$\times 10^{-12}$ & 0.77 (16.1/21)  \\
IIc & $0.41^{*}$& $0.41^{*}$& 0.25$^{*}$& $0.82^{+0.42}_{-0.42}$ & $0.27^{+0.07}_{-0.11}$ & 2.2$\times 10^{-12}$ & 2.8$\times 10^{-12}$ & 0.78 (16.4/21) \\
            \noalign{\smallskip}
            \hline
\noalign{\smallskip}
\multicolumn{9}{p{15.5cm}}{\scriptsize
$^*$ Fixed value;
$^{\mathrm{a}}$ I: (R\&S + PL$^{\rm abs}$)$^{\rm abs}$, II: (M\&K + PL$^{\rm abs}$)$^{\rm abs}$, (R\&S$^{\rm abs}$: Absorbed Raymond/Smith plasma, M\&K$^{\rm abs}$: Absorbed Mewe/Kaastra plasma, PL$^{(\rm abs)}$: (Absorbed) Power--law;
$^{\mathrm{b}}$ Column density of the diffuse absorber;
$^{\mathrm{c}}$ Metallicity (referring to the solar value);
$^{\mathrm{d}}$ Column density of the compact absorber;
$^{\mathrm{e}}$ Photon index;
$^{\mathrm{f}}$ Absorbed X-ray flux between 0.2\,keV and 8.0\,keV;
$^{\mathrm{g}}$ Unabsorbed X-ray flux between 0.2\,keV and 8.0\,keV, i.e., $N^{\mathrm{(b)}}_{\rm H(gal.)}$ and $N^{\mathrm{(d)}}_{\rm H(torus)}$ are set to 0
}
\end{tabular*}
}
\end{table*}


\section{Discussion}
\label{summ}
The {\it CHANDRA} data provide for the first time direct evidence for
{ jet-associated X-ray emission in NGC\,1052.}
The diffuse, extended X-ray emission can be best approximated with a thermal plasma model
with { $kT \sim 0.4-0.5$\,keV}. This temperature is consistent with the thermal
component found earlier by Weaver et al. (1999) using {\it ASCA} and {\it ROSAT} data.
Its absorbed flux is only { $\sim 3$\%} of the nuclear X-ray emission { but the intrinsic
(absorption corrected) extended emission might contribute up to 14\% to the
total unabsorbed X-ray flux of NGC\,1052.} 
Because of the { considerable} pile-up degradation of
the {\it CHANDRA} data, no firm conclusions on the photon index of the nucleus spectrum can
be deduced.
The derived column density of hydrogen towards the compact X-ray core
(depending on the applied model) { of $0.5-0.8 \times 10^{22}$\,cm$^{-2}$ is in good
agreement with the absorbing column density of ionized material towards the 
VLBI-jet derived by Kadler et al. (\cite{Kad02}) and Kadler et al. (in prep.). 
This suggests that the
nuclear X-ray emission of NGC\,1052 might be produced on the same scales
as the parsec-scale structures imaged by VLBI at high frequencies.}

The detection of a diffuse region of
X-ray emitting gas with a thermal spectrum and the same extent 
as the kiloparsec-scale radio jet suggests that 
jet-triggered shocks might play an important role in NGC\,1052. 
In such a model the kinetic power of the radio jet is 
partially
converted into X-ray emission. 
The optical morphology in the H$\alpha$ filter substantiates
this
picture as was noted earlier by Allen et al. (\cite{All99}). The alignment of
the radio jet and the optical emission cone visible in
Fig.~\ref{fig:structure_map} implies that the ionization cone
might be drilled out by the radio jets,  resulting in a
predominantly shock-excited, conical
narrow-line region (see e.g., Dopita \cite{Dop02}).
Shocks might occur also on larger scales giving rise to the
soft thermal X-ray emission associated with the radio jet/lobe structure
in NGC\,1052.
A rough estimate (see Kadler et al. \cite{Kad03a})
shows that the soft thermal
X-ray spectrum associated with the radio jet of NGC\,1052 can be explained
in terms of the kinetic jet power being partially converted into X-ray emission
originating in shocks driven into the ambient medium.
(A more detailed model of the relation between jet-driven shock-activity
and the spectral shape of the extended X-ray emission in NGC\,1052
will be discussed in a forthcoming paper.)
The comparison of the large-scale 
distribution of radio emission in NGC\,1052 between two epochs separated by
$\sim$ 15 years indeed shows activity on { kiloparsec-scales}. This
substantiates the idea that shocks in the interstellar medium form at the
working surfaces of active regions (hotspots and knots).
Moreover, recent numerical simulations (e.g., Zanni et al. \cite{Zan03}) show
that jets in radio galaxies can inflate over-pressured cocoons that drive shocks
into the ambient gas resulting in morphologies (in the case of weak shocks)
very similar to what is observed in NGC\,1052: a cavity of hot X-ray emitting
gas in conjunction with a local deficit of X-ray emission around the hotspots. 
A deeper {\it CHANDRA} observation with an improved  photon statistic { compared to} the
observation discussed here { would} provide both a higher sensitivity to the weak
diffuse emission and a higher resolution. 
Additionally, the full resolution of {\it CHANDRA} of
$\sim 0.5$\,arcsec would allow one to study in more detail the connection between the
knots in the diffuse X-ray emission and the optical emission knots.

\begin{acknowledgements}
We thank G.\,V.\,Bicknell for helpful discussions and important suggestions.
This research used the {\it CHANDRA} Data Archive (CDA) which is part of the {\it CHANDRA} X-ray Observatory Science Center (CXC) 
which is operated for NASA by the Smithsonian Astrophysical Observatory. 
We used {\it CHANDRA} data from an experiment
planned and scheduled by G. P. Garmire.
We made use of the data archive at the Space Telescope Science Institute
which is operated by the Association of Universities for Research
in Astronomy, Inc. under NASA contract NAS 5-26555.
MERLIN is a National Facility operated by the University of Manchester at Jodrell Bank Observatory 
on behalf of PPARC. We made use of public MERLIN data from an experiment planned and scheduled by
A.\,Pedlar.
This research has made use of NASA's Astrophysics Data System.
\end{acknowledgements}
\bibliographystyle{aa}

\end{document}